\documentclass[%
 amsmath,amssymb,
 aps,
]{revtex4-2}

\usepackage{amsmath,amssymb}
\usepackage{mathrsfs} 
\usepackage{xcolor}

\usepackage{siunitx}
\usepackage{mathptmx}
\usepackage{graphicx}

 \newcommand{\bfE}{\mathbf{E}}
\newcommand{\bE}{\mathbf{E}}
\newcommand{\bfB}{\mathbf{B}}

\newcommand{\bB}{\mathbf{B}}
\newcommand{\bfJ}{\mathbf{J}}
\newcommand{\bJ}{\mathbf{J}}
\newcommand{\bfv}{\mathbf{v}}

\newcommand{\bfx}{\mathbf{x}}

\newcommand{\bbS}{\mathbb{S}}

\begin{document}

\title{Advances in the Implementation of the Exactly Energy Conserving Semi-Implicit  (ECsim) Particle in Cell Method}


\email{giovanni.lapenta@kuleuven.be}

\author{Giovanni Lapenta}
\affiliation{Center for mathematical Plasma Astrophysics, Department of Mathematics, University of Leuven, KULeuven, Belgium} 




\begin{abstract}
The Energy Conserving semi-implicit method (ECsim), presented by Lapenta in 2017, is a Particle in Cell (PIC) algorithm for the simulation of plasmas.  Energy conservation is achieved within a semi-implicit formulation that does not require any non-linear solver. A mass matrix is introduced to express linearly the particle-field coupling. With the mass matrix the algorithm preserves energy conservation to machine precision. The construction of the mass matrix is the central nature of the method and also the main cost of the computational cycle. We analyze here three methods that modify the construction of the mass matrix. First, we consider how the sub-cycling of the particle motion modifies the mass matrix. Second, we introduce a form of smoothing that reduces the noise while retaining exact energy conservation. Finally, we discuss an approximation of the mass matrix that transform the ECsim scheme to the implicit moment method.
\end{abstract}

\maketitle

\section{Introduction}

The Energy Conserving semi implicit method (ECSim) is an algorithm for plasma simulation based on the particle in cell (PIC) approach \citep{lapenta2017exactly}. PIC methods can be explicit, semi-implicit or fully implicit. Plasmas are governed by two sets of equations: the equations for the motion of the particles and the equations for the evolution of the fields. The two sets of equations are coupled because the field equations need the sources (current and charge) from the particles and the particles need the fields to compute the force. As a particle moves, the fields are modified and as the fields change the forces on the particles are modified. This link is central to the physics of plasmas: plasmas are collective sets of particles interacting via the fields. The coupling between particles and fields is non-linear and to represent it in discretized equations in its fullness one needs fully implicit methods. In fully implicit methods, the particle equations of motion and the field equations are solved together within a non-linear solver, such as the Newton-Krylov approach \citep{markidis2011energy,chen2011energy}. In explicit methods, conversely, the coupling between particles and fields is suspended for a small time step~\citep{birdsall-langdon}. In that small time interval one assumes that the known fields can be used unchanged for moving the particles and the particle information can be used unchanged to evolve the fields. This has three major consequences. 

First, the explicit method is very simple, no iteration is needed and explicit PIC can be implemented as some of the most efficient algorithm known in computer science, consistently being a top performance achiever on any new computer architecture introduced. For example, PIC was one of the first applications to reach petascale performance~\citep{vpic}. Implicit PIC is much more complex in its implementation. Especially on massively parallel computers, reaching high efficiency is a challenge.

Second, in explicit PIC, the time step becomes limited by numerical stability considerations, requiring to use high resolution. The peculiarity of PIC is that the resolution needs not just be refined in time, to resolve the electron plasma frequency $\omega_{pe}\Delta t<2$ \citep{hockney-eastwood}, but also in space, to avoid the so-called finite-grid instability~\citep{birdsall-langdon}. This limitation is removed by the implicit approach that allows one to select grid spacing and time step based on the accuracy needed and not on the stability of the numerical algorithm~\citep{lapenta2012particle}. 

Third, in explicit PIC, energy is not conserved. Using a good resolution, energy is acceptably maintained. There is a secular trend of energy increase always\citep{birdsall-langdon}, but as the resolution is relaxed closer to the stability limit of the finite grid instability, the energy increase becomes more severe, until at the instability limit it starts to grow exponentially. This effect cannot be avoided but it can be improved by using smoothing and higher order interpolation techniques. Recent  structure-preserving geometric particle-in-cell methods use  symplectic integrators  to ensure local energy conservation at small time steps \citep{jianyuan2018structure}. The implicit PIC method, instead, conserves energy exactly, whatever resolution is used~\citep{markidis2011energy,chen2011energy}. This feature is physically important and practically impactful. Physically, of course, confidence comes form using an algorithm that preserves one of the most established properties in physics: conservation of energy. If energy starts to spontaneously increase confidence in the results is shaken. Practically, lack of energy conservation requires a tedious and careful tuning of the parameters to make sure the simulation does not increase its energy excessively, leading in some situations to excessively resolved models that need to use much more resolution than the processes of interest require.

The semi-implicit PIC method tries to make a compromise and retain some of the advantages of both approaches. In semi-implicit methods the particles and the fields are still advanced together and an iteration is needed but the coupling is linearized and the iteration uses linear solvers.  Different methods are used to linearize the coupling. The implicit moment method formulates the particle response to changes in the fields using the moment closure method~\citep{brackbill-forslund}. The direct implicit method uses a formal linear expansion of the coupling operator~\citep{directimplicit}. In all these approaches the stability properties of the semi-implicit method are superior to explicit methods and allow a good compromise in the resolution needed~\citep{lapenta2012particle}. However, energy is not conserved. Unlike explicit PIC, energy can either increase or decrease depending on the implementation because dissipation terms are included in the algorithm to suppress energy growth. 

The ECsim approach is the first semi-implicit PIC to retain exact energy conservation as in the fully implicit PIC. ECsim uses a mathematical construct called mass matrix to express the coupling between particles and fields. With the mass matrix, the coupling is linear but energy is conserved exactly. Below we review how this was achieved~\citep{lapenta2017exactly}. 

Since its recent  introduction, ECsim has found application in a number of applications in space~\citep{lapenta2017multiple,walker2019embedding,zhou2019embedded,lapenta2020multiscale,lapenta2022we} 
and in fusion~\citep{gonzalez2019ecsim,park2019discovery}. An important improvement has removed the lack of charge conservation in the original scheme \citep{chen2019gauss,pinto2022semi}. We report here some extensions of the method that can widen its practical applicability. 

First, we describe a method to introduce smoothing to reduce noise, while retaining exact energy conservation. Smoothing is an affordable way to introduce effectively a higher order interpolation scheme. It removes the high frequency part of the spectrum. Without special attention, smoothing will tend to break energy conservation by just removing high frequency fluctuations from the system. We present, instead, a method that conserves energy.

Second, sub-cycling is a convenient approach in plasma simulation to address the faster scales seen by the particles. In some applications, the particle response (or that of a subset of the particle population) is faster than the evolution of the fields and it is beneficial to move the particles several times without advancing the fields. In explicit PIC, the operation is very simple since fields and particles are not advanced together. In implicit and semi-implicit PIC, instead,  sub-cycling needs to be done also in the coupling requiring to modify the algorithm to compute the current used in the field solution. 

Finally, we illustrate how the mass matrix formulation opens up the opportunity to approximate the mass matrix in certain limit cases to reduce the cost of the simulation. 

The paper is organizes as follows. Section \ref{summary} recaps the key properties of the ECsim needed for the discussion. Section \ref{smoothing} introduces how  smoothing can be implemented to reduce  noise while retaining the property of energy conservation. Section \ref{sub-cycling} is dedicated to the algorithm for  sub-cycling the particle motion within the ECsim scheme. Section \ref{simplification} derives a limit case of the mass matrix formulation that transforms the ECsim algorithm into the standard implicit moment method~\citep{brackbill-forslund}. Results are presented in Sect.~\ref{results} and final conclusions are drawn in Sect.~\ref{discussion}.

\section{Summary of the Energy Conserving semi-implicit (ECsim) method}
\label{summary}
The energy conserving semi-implicit method~\citep{lapenta2017exactly} is based on a formulation of the mover similar to the classic leap-frog scheme:

\begin{equation}
\begin{array}{c}
\displaystyle \bfx_p^{n+1/2}=\bfx_p^{n-1/2} + \Delta t \bfv_p^{n}\\ \\
\displaystyle \bfv_p^{n+1}=\bfv_p^{n} + \frac{q_p \Delta t}{m_p} \left(  \bfE^{n+\theta}(\bfx_p^{n+1/2}) + \overline{\bfv}_p\times \bfB^{n}(\bfx_p^{n+1/2}) \right)
\end{array}
\label{thetaECSIM}
\end{equation}
where $\overline{\bfv}_p= (\bfv_p^{n+1}+ \bfv_p^{n})/2$ and $\bfE^{n+\theta} = \theta \bfE^{n+1} + (1-\theta)  \bfE^{n}$. 

This mover differs from both the explicit leap-frog mover~\cite{boris1970relativistic} and the implicit $\theta$-mover \citep{Brackbill:1985}: it combines the first equation from the explicit mover with the second equation from the implicit mover but with an important difference: the electric and magnetic fields are computed at the known position $\bfx_p^{n+1/2}$ rather than at the unknown position $\overline{\bfx}_p$. The important consequence is that the particle equations can be solved directly without any iteration needed among themselves. Instead in the standard $\theta$-mover, a predictor-corrector iteration is required. Yet the mover is still implicit because the new fields are not known until the field equations are solved. The ECsim method retains the coupling between advanced fields and advanced particles, requiring the solution of a linear coupled system. However, the mover itself does not require any iteration, a substantial simplification. 

In  the ECsim scheme, the electric and magnetic fields are computed at the known position $\bfx_p^{n+1/2}$. In the  $\theta$-scheme, instead, they are computed at the unknown position $\overline{\bfx}_p$. These two positions are conceptually the same, they express the particle position at the mid-time between the old and new evaluations of the velocity. But one is computed explicitly, in the leap-frog sense, while the other is computed as part of a predictor-corrector iteration \cite{vu1995accurate,lapenta05}. Both methods are second order accurate but the ECsim scheme is simpler to compute. This simplicity is not just a virtue in itself but leads to an important consequence: the simplicity allows us to formulate the coupling with the fields in a way that insures exact energy conservation without requiring non-linear iterations. The  $\theta$-scheme can be made energy conserving but at the cost of a fully non-linear iteration requiring a non-linear solver \citep{markidis2011energy,chen2011energy}.The ECsim scheme allows exact energy conservation without requiring any non-linear iteration. 

The properties of stability are determined by the field-particle coupling and in this sense the method is still implicit. For this reason, not requiring any non-linear iteration but still requiring a liner solver to deal with the field-particle coupling, the method is semi-implicit. This nomenclature is to distinguish it from the fully implicit method that requires the non-linear iteration. 

Note that the force term is written using the magnetic field at the initial time level $\bfB^{n}(\bfx_p^{n+1/2})$ but the electric field is written at the advanced intermediate level $\bfE^{n+\theta}(\bfx_p^{n+1/2})$. The reason for this choice is simplicity and the fact that the magnetic field does no work and using the old time level does not introduce any loss of energy conservation.

The coupling of particles and fields  require to interpolate the fields to  the particle positions:
\begin{align}
\bfE_p^{n+\theta}&=\bfE^{n+\theta}(\bfx_p^{n+1/2})=\sum_g \bfE_g^{n+\theta} W(\bfx_p^{n+1/2}-\bfx_g)\\
\bfB_p^{n}&=\bfB^{n+\theta}(\bfx_p^{n+1/2})=\sum_g \bfB_g^{n} W(\bfx_p^{n+1/2}-\bfx_g)
\label{interpE}
\end{align}
We used a generic index $g$ for the grid. In the specific implementation within the iPic3D code~\citep{ipic3d, gonzalez2018performance} the electric field and magnetic field are not colocated and $g$ label either centers (for $\bfB$) or vertices (for $\bfE$). 
Here we simplify the notation as: $\bfB_p^{n}=\bfB^{n}(\bfx_p^{n+1/2})$ and $\bfE_p^{n+\theta}=\bfE^{n+\theta}(\bfx_p^{n+1/2})$. In our implementation, the interpolation function $W$ is a b-splines of order $\ell=1$ \cite{bspline}:
\begin{equation}
W(\bfx_p-\bfx_g)=b_\ell(x_p-x_g)b_\ell(y_p-y_g)b_\ell(z_p-z_g)
\end{equation}
This expression reduces trivially in 1D for the examples reported below.

For the Maxwell's equation we use the standard $\theta$-scheme \cite{ipic3d}:
\begin{equation}
\begin{array}{ccc}
\displaystyle \nabla_g \times \bfE^{n+\theta} + \frac{1}{c} \frac{\bfB^{n+1}_g-\bfB^n_g}{\Delta t} =0\\ \\
\displaystyle \nabla_g \times \bfB^{n+\theta} - \frac{1}{c} \frac{\bfE^{n+1}_g-\bfE^n_g}{\Delta t} =\frac{4\pi}{c} \overline{\bfJ}_g
\end{array}
\label{maxwell-discrete}
\end{equation}
The spatial operators in eq. (\ref{maxwell-discrete}) are discretized on the  grid labelled by $g$ introduced above. 


The coupling of the field equations with the particles is expressed by the current for each species:
\begin{equation}
\overline{\bfJ}_{sg}=\frac{1}{V_g}\sum_{p \in s} q_p \overline{\bfv}_p W(\bfx_p^{n+1/2}-\bfx_g)
\label{currentECSIM}
\end{equation}
where the summation is over the particles of the same species, labeled by $s$.

As with the $\theta$-mover, the velocity equation can be rewritten in the equivalent  form~\citep{vu}: 
\begin{equation}
\overline{\bfv}_p=\widehat{\bfv}_p+
\beta_s\widehat{\bE}_p
\label{theta-rotated}
\end{equation}
with:
\begin{equation}
\begin{array}{c}
 \widehat{\bfv}_p = {\alpha}^n_p  \bfv^n_p \\ \\
\widehat{\bE}_p = {\alpha}^n_p 
\bE_p^{n+\theta} 
\end{array}
\label{hatted}
\end{equation}
and the  rotation matrix ${\alpha}_p^n$ 
given by:
\begin{equation}
{\alpha}_p^n =  \frac{1}{1+(\beta_s B_p^{n})^2}
\left(\mathbb{I}-\beta_s \mathbb{I} \times \bB_p^n +\beta_s^2
\bB_p^n \bB_p^n \right)
\label{alpha}
\end{equation}
where $\mathbb{I}$ is the dyadic tensor (matrix with diagonal of 1) and $\beta_s=q_p \Delta t/2m_p$ (independent of the particle weight
and unique to a given species).  The elements of the rotation matrix are indicated as 
${\alpha}^{ij,n}_p $ with label $i$ and $j$ referring to the 3 components of the vector space ($x$, $y$, $z$).  

Substituting then eq.~(\ref{theta-rotated}) into eq.~(\ref{currentECSIM}), we obtain without any approximation or linearization:
\begin{equation}
\overline{\bfJ}_{sg}=\frac{1}{V_g}\sum_p q_p \widehat{\bfv}_p W_{pg} +\frac{\beta_s}{V_g}\sum_p q_p  \widehat{\bfE}_p^{n+\theta} W_{pg}
\label{currentECSIM1}
\end{equation}
where we shortened the notation $W_{pg}=W(\bfx_p^{n+1/2}-\bfx_g)$  and the summation is intended over all particles of species $s$.

Using  eq.~(\ref{hatted}), the  expression for the current becomes:
 \begin{equation}
\overline{\bfJ}_{sg}=\widehat{\bJ}_{sg}+\frac{\beta_s}{V_g}\sum_p q_p {\alpha}^{n}_p
\bE_p^{n+\theta}  W_{pg}
\label{currentECSIM2}
\end{equation}
where we defined: 
\begin{equation}
\widehat{\bJ}_{sg} = \frac{1}{V_g}\sum_p q_p  \widehat{\bfv}_p W_{pg}
\label{hattedmoments}
\end{equation} 

Computing then the electric field on the particles by interpolation form the grid as in eq. (\ref{interpE}), it follows that:
\begin{equation}
\overline{\bfJ}_{sg}=\widehat{\bJ}_{sg}+\frac{\beta_s }{V_g}\sum_p \sum_{g^\prime} q_p {\alpha}^{n}_p
 \bE_{g^\prime}^{n+\theta}  W_{pg^\prime} W_{pg}
\label{currentECSIM3}
\end{equation}
Exchanging the order of summation we obtain:
\begin{equation}
\overline{\bfJ}_{sg}=\widehat{\bJ}_{sg}+\frac{\beta_s }{V_g}\sum_{g^\prime} M_{s,gg^\prime}
 \bE_{g^\prime}^{n+\theta}  
\label{currentECSIMfinal}
\end{equation}
where we have defined the actor in the leading role of the ECsim scheme: the mass matrix~\cite{burgess1992mass}:
\begin{equation}
M_{s,gg^\prime}^{ij} = \sum_p q_p {\alpha}^{ij,n}_p W_{pg^\prime} W_{pg}
\label{mass-matrices}
\end{equation}
There are $3v$ (where $v$ is the number of velocity directions) mass matrices and in matrix notation they can be written as $M_{gg^\prime}$, that is without the indices $i,j$ for the vector directions.

The mass matrices $M_{s,gg^\prime}$ that are the most important aspect of the ECsim method and are also the most expensive part of the computation~\citep{gonzalez2018performance}. A number of symmetries can be used to reduce the cost. Speed up of the construction can be achieved using offloading to accelerator processors (e.g. graphical processing units, GPU)~\citep{boella2022adding}. The mass matrices, eq. (\ref{currentECSIMfinal}), provide  an explicit linear link between the advanced current at the mid-point of the time step and the electric field at the advanced time. This linear relationship can be substituted into the discretized Maxwell's equations (\ref{maxwell-discrete}) to form a linear set of equations:
\begin{equation}
\left\{ \begin{array}{l}
\displaystyle \nabla_g \times \bfE^{n+\theta} + \frac{1}{c} \frac{\bfB^{n+1}-\bfB^n}{\Delta t} =0\\ \\
\displaystyle \nabla_g \times \bfB^{n+\theta} - \frac{1}{c} \frac{\bfE^{n+1}-\bfE^n}{\Delta t} =\frac{4\pi}{c} \left( \widehat{\bJ}_{g}+\sum_{g^\prime} M_{gg^\prime}
 \bE_{g^\prime}^{n+\theta} \right)
\end{array}
\right.
\label{maxwellECsim}
\end{equation}
where the total current is $\widehat{\bJ}_{g} = \sum_s \widehat{\bJ}_{sg}$ and the species summed mass matrices, that written by elements are:
\begin{equation}
M_{gg^\prime}^{ij} =\sum_s\frac{\beta_s }{V_g}M_{s,gg^\prime}^{ij}
\end{equation}

The direct link provided by the mass matrix is analytically exact for the original set of discretized equations. Unlike the implicit moment method where the equations have to be approximated by Taylor series expansion~\cite{vu}, here the link is still exactly the same as in the original set of discretized equations. Having eliminated the need for any approximation or Taylor series expansion is the reason why ECsim conserves energy exactly. 

We consider now how energy conservation can be shown in the case  $\theta=1/2$. It is important for the derivations below to consider what key steps enable energy conservation.  The inner product of  the velocity equation (\ref{thetaECSIM}) with the average speed, $\overline{\bfv}_p$ gives by summing over all particles:
\begin{equation}
\frac{1}{2}\sum_p \left( m_p (\bfv_p^{n+1})^2 - (\bfv_p^{n})^2 \right) =  \Delta t  \sum_p  \left ( q_p  \sum_g \overline{\bfv}_p \cdot\overline{ \bfE}_g W_{pg} \right)
\end{equation}
where the electric field is computed as average consistent with the choice $\theta=1/2$ and the magnetic field drops out as obvious from the properties of the cross product. Exchanging the summation over particles and cells leads to:
\begin{equation}
\frac{1}{2}\sum_p \left( m_p (\bfv_p^{n+1})^2 - (\bfv_p^{n})^2 \right) =  \Delta t  \sum_g  \overline{ \bfJ}_g \cdot \overline{ \bfE}_g 
\label{energy-particles}
\end{equation}
where it is recognized that $ \overline{ \bfJ}_g= \sum_p q_p  \overline{\bfv}_p W_{pg} $.

Multiplying the first equation  (\ref{maxwell-discrete}) by $\overline{\bfB}_g$ and the second by $\overline{\bfE}_g$ and summing them leads to:
\begin{equation}
\begin{array}{l}
\displaystyle\frac{ (\bfB_g^{n+1})^2-(\bfB_g^n)^2 }{2c}+\frac{ (\bfE_g^{n+1})^2-(\bfE_g^n)^2 }{2c} = \\ \\
\displaystyle \Delta t \left( \frac{4\pi}{c} \overline{\bfJ}_g \cdot \overline{\bfE}_g + \bfE_g \cdot \nabla_g \times \bfB - \bfB_g \cdot \nabla_g \times \bfE \right)
\end{array}
\end{equation}
Assuming a mimetic grid discretization that preserves the continuum properties of the operators and summing over all grid points gives:
\begin{equation}
\begin{array}{l}
\displaystyle\sum_g \frac{ (\bfB_g^{n+1})^2-(\bfB_g^n)^2 }{4 \pi}+ \sum_g\frac{ (\bfE_g^{n+1})^2-(\bfE_g^n)^2 }{4\pi} = \\ \\
\displaystyle \Delta t \sum_g\overline{\bfJ}_g \cdot \overline{\bfE}_g +
\frac{c\Delta t}{4 \pi}\sum_g \nabla_g \cdot  (\bfE_g \times \bfB_g )
\end{array}
\label{energy-fields}
\end{equation}

This conservation law states that the  variation of the magnetic and electric energy, as measured on the grid, equals the amount exchanged  with the particles and carried by the grid-discretized divergence of the Poynting flux.  For energy to be conserved in the system, the energy exchange term on the particle equations (eq. (\ref{energy-particles}))  needs to be identical to that on the field equations, (eq. (\ref{energy-fields})). This term is  indeed identically equal to $\sum_g\overline{\bfJ}_g \cdot \overline{\bfE}_g $ in both equations. Energy conservation is enforced exactly, to round off.  

Besides guaranteeing physical conservation of energy, a cornerstone in any physical model, the existence of this conservation constraint also guarantees a form of non-linear stability of the discretized equations~\citep{lapenta2017exactly}  expanding the stability of the semi-implicit method compared with the moment implicit scheme~\citep{lapenta2017multiple,gonzalez2019ecsim}.

\section{Smoothing with the mass matrix formulation}
\label{smoothing}
Smoothing can be designed to be compatible with the energy conserving properties of the mass matrix.
We choose to smooth only the electric field, since the magnetic field tends to be much smoother in PIC simulations and smoothing is not needed. 

From the proof of energy conservation recapped above, it is clear that for energy conservation the smoothing of the current must be done in the same way as that of the electric field in the mover. 
Starting from the mover and calling $\bbS_{gg^\prime}$ the smoothing operator, we define a smoothed electric field  on the grid as:
\begin{equation}
\bfE^{SM}_{g} =\sum_{g^\prime}\bbS_{gg^\prime} \bfE^{n+\theta}_{g^\prime}
\label{interpsm}
\end{equation}
From eq, (\ref{interpsm}),  the smoothed electric field acting on a particle can be computed as:
\begin{equation}
\bfE^{SM}_p=\sum_g \bfE^{SM}_g W(\bfx_p^{n+1/2}-\bfx_g)
\label{interpEsm}
\end{equation}

The second equation of motion,   eq. (\ref{thetaECSIM}), then uses the smoothed electric field as:
\begin{equation}
\bfv_p^{n+1}=\bfv_p^{n} + \frac{q_p \Delta t}{m_p} \left(  \bfE^{SM}_p + \overline{\bfv}_p\times \bfB^{n}_p \right)
\label{interpeom}
\end{equation}
where $\bfB^{n}_p$ is still computed as above.

From eq. (\ref{interpeom}), we can compute again the current as
\begin{equation}
\overline{\bfJ}_{sg}=\widehat{\bJ}_{sg}+\frac{\beta_s }{V_g}\sum_{g^\prime} M_{s,gg^\prime}
 \bE^{SM}_{g^\prime}  
\label{currentECSIMfinalsm}
\end{equation}
The energy exchange term for the particles then becomes
\begin{equation}
\frac{1}{2}\sum_p \left( m_p (\bfv_p^{n+1})^2 - (\bfv_p^{n})^2 \right) =  \Delta t \sum_s \sum_g \left( \widehat{\bJ}_{sg}+\frac{\beta_s }{V_g}\sum_{g^\prime} M_{s,gg^\prime}
 \bE^{SM}_{g^\prime} \right) \cdot  \bfE^{SM}_g 
\label{energy-particles-sm}
\end{equation}

Applying now smoothing to the source term of the second of the Maxwell eq. (\ref{maxwellECsim}), we have:
\begin{equation}
 \nabla_{g} \times \bfB^{n+\theta} - \frac{1}{c} \frac{\bfE^{n+1}_{g} - \bfE^n_{g} }{\Delta t} =\frac{4\pi}{c}  \sum_s \sum_{g^\prime} \bbS_{gg^\prime}\left(\widehat{\bJ}_{sg^\prime}+\frac{\beta_s }{V_g}\sum_{g^{\prime\prime}} M_{s,g^\prime g^{\prime\prime}}
 \bE^{SM}_{g^{\prime\prime}}  \right) 
\end{equation}
where the last term is expressed from eq. (\ref{currentECSIMfinalsm}). 

For the fields, the energy integral then becomes:
\begin{equation}
\begin{array}{l}
\displaystyle\sum_g \frac{ (\bfB_g^{n+1})^2-(\bfB_g^n)^2 }{4 \pi}+ \sum_g\frac{ (\bfE_g^{n+1})^2-(\bfE_g^n)^2 }{4\pi} -
\frac{c\Delta t}{4 \pi}\sum_g \nabla_g \cdot  (\bfE_g \times \bfB_g )= \\ \\
\displaystyle \Delta t \sum_s \sum_g\sum_{g^\prime}\bbS_{gg^\prime} \left(\widehat{\bJ}_{sg^\prime}+\frac{\beta_s }{V_{g^\prime}}\sum_{g^{\prime\prime}} M_{s,g^\prime g^{\prime\prime}}
 \bE^{SM}_{g^{\prime\prime}}  \right)  \cdot \overline{\bfE}_g 
\end{array}
\label{energy-fields-sm}
\end{equation}

For the two energy integrals to be the same the exchange term seen by the particles must be equal to that seen by the fields. The right-hand sides of eq.(\ref{energy-particles-sm}) must then equal that of eq.
(\ref{energy-fields-sm}):
\begin{equation}
\begin{split}
\sum_g\sum_{g^\prime}\bbS_{gg^\prime} \left(\widehat{\bJ}_{sg^\prime}+\frac{\beta_s }{V_{g^\prime}}\sum_{g^{\prime\prime}} M_{s,g^\prime g^{\prime\prime}}
 \bE^{SM}_{g^{\prime\prime}}  \right)  \cdot \overline{\bfE}_g  = \\ \sum_g \left( \widehat{\bJ}_{sg}+\frac{\beta_s }{V_g}\sum_{g^{\prime\prime}} M_{s,gg^{\prime\prime}}
 \bE^{SM}_{g^{\prime\prime}} \right) \cdot  \sum_{g^\prime}\bbS_{gg^\prime} \overline{\bfE}_{g^\prime} 
 \end{split}
 \end{equation}
 Switching $g$ with $g^\prime$ (just names) in the right-hand side, the equivalence above holds  when the smoothing operator is symmetric (i.e. the matrix repenting it is symmetric), a common property shared by many smoothing operators \cite{chen2011energy}. 
 
Note that we smooth the electric field but not the magnetic field that tends to be less noisy by its nature. 
 
\section{Sub-cycling with the mass matrix formulation}
\label{sub-cycling}

A mass matrix can be defined also in presence of sub-cycling or orbit averaging movers. The velocity update of ECsim can be reformulated for sub-cycling as:
\begin{equation}
\displaystyle \bfv_p^{\nu+1}=\bfv_p^{\nu} + \frac{q_p\Delta t_\nu}{m_p} \left(  \bfE^{n+\theta}(\bfx_p^\nu) + \frac{\bfv_p^{\nu+1}+\bfv_p^{\nu}}{2}\times \bfB^{n}(\bfx_p^{\nu}) \right)
\label{sub-cycling}
\end{equation}
We  assume that the time step $\Delta t$ between field updates is subdivided into  $N_\nu$ not necessarily equal sub-steps $\Delta t_\nu$. 

The positions for the field evaluations, $\bfx_p^\nu$,  during the sub-cycle can be computed in different ways. The simplest is to assume a straight orbit within $\Delta t$, similar to  the leap-frog approach: 
\begin{equation}
    \bfx_p^\nu = \bfx_p^{n-1/2} + \bfv_p^{n}\sum_{\nu^\prime=0
    }^{\nu^\prime=\nu}\Delta t_{\nu^\prime}  
\end{equation}
that can all be computed at once since the same velocity is used for all points along the trajectory. 
The first step starts from the old position: $\bfx_p^{\nu=0}= \bfx_x^{n}$ and old velocity $\bfv_p^{\nu=0}= \bfv_x^{n}$ and the last step leads to the final position $\bfx_p^{\nu=N_\nu}= \bfx_p^{n+1}$ and final velocity $\bfv_p^{\nu=N_\nu}= \bfv_p^{n+1}$ .  The fields are assumed to be those computed at the time level $\theta$ within the field update time step $\Delta t$. 

Another promising  approach is to recall that most often in plasma physics particles are not moving in straight lines but rather they are frozen into the field lines, moving in cyclotron orbits with drifts due to the in-homogeneity of the fields. In the spirit of gyro-averaging, certain applications of  implicit method might need to step over the gyration time scale and the positions of the particles used in eq.(\ref{sub-cycling}) would then be chosen to achieve accurate gyro-averaging, for example taking $N_\nu$ positions along the gyro-orbit of a particle \cite{lee1987gyrokinetic} to compute an average force on the particle's center of gyration.

The example of the two strategies above for computing  the intermediate positions can be made in a single explicit step that generates all positions at once: in this case each substep contribution to the mass matrix and the moments can be computed in parallel, greatly improving the parallel performance. In practice, the $N_\nu$ operations required by the substepping algorithm can all be done in parallel in an embarrassingly parallel approach that scales ideally on supercomputers: no communication between the particles and between the substeps is needed.

The equation (\ref{sub-cycling}) can be inverted with the same  vector manipulations used for eq. (\ref{thetaECSIM}), to obtain : 
\begin{equation}
\frac{\bfv_p^{\nu+1}+ \bfv_p^{\nu}}{2}=  \widehat{\bfv}_p^{\nu}+
\beta_s\widehat{\bE}_p^\nu
\label{theta-rotated-sub}
\end{equation}
where hatted quantities have been rotated by the magnetic field computed at the location $\bfx_p^{\nu}$:
\begin{equation}
\begin{array}{c}
 \widehat{\bfv}_p^{\nu} = {\alpha}^\nu_p  \bfv^\nu_p \\ \\
\widehat{\bE}_p^\nu = {\alpha}^\nu_p 
\bfE^{n+\theta}(\bfx_p^{\nu}) 
\end{array}
\label{hatted-sub}
\end{equation}
via a rotation matrix ${\alpha}_p^n$ 
defined as in the case of a singe step but the magnetic field computed at the last substep position:
\begin{equation}
{\alpha}_p^\nu =  \frac{1}{1+(\beta_s B^{n}(\bfx_p^{\nu}))^2}
\left(\mathbb{I}-\beta_s \mathbb{I} \times \bfB^{n}(\bfx_p^{\nu}) +\beta_s^2
\bfB^{n}(\bfx_p^{\nu}) \bfB^{n}(\bfx_p^{\nu}) \right)
\label{alpha-sub}
\end{equation}
From eq.~(\ref{theta-rotated-sub})  we  can obtain directly from its definition (\ref{currentECSIM}) the mean current over all sub-steps, without any further approximation or linearization:
\begin{equation}
\overline{\bfJ}_{sg}=\frac{1}{V_g}\sum_p q_p  \sum_\nu \frac{\Delta t_\nu}{\Delta t}\left( \widehat{\bfv}_p^\nu + \beta_s  \widehat{\bfE}_p^{\nu} \right ) W_{pg}^\nu
\label{currentECSIM1-sub}
\end{equation}
where $W_{pg}^\nu=W(\bfx_p^{\nu}-\bfx_g)$.
Using now the definitions of the hatted quantities, eq. (\ref{hatted-sub}), we can cast the mean current in the same form as in the single step formulation:
\begin{equation}
\overline{\bfJ}_{sg}=\widehat{\bJ}_{sg}+\frac{\beta_s }{V_g}\sum_{g^\prime} M_{s,gg^\prime}
 \bE_{g^\prime}^{n+\theta}  
\label{currentECSIMfinal-sub}
\end{equation}
but with the new definition for 
\begin{equation}
\widehat{\bJ}_{sg} = \frac{1}{V_g}\sum_p q_p  \sum_\nu \frac{\Delta t_\nu}{\Delta t} \widehat{\bfv}_p^\nu W_{pg}^\nu
\label{hattedmoments-sub}
\end{equation}
and the mass matrix for a sub-cycled trajectory defined by:
\begin{equation}
M_{s,gg^\prime}^{ij} = \sum_p q_p \sum_\nu \frac{\Delta t_\nu}{\Delta t} {\alpha}^{ij,\nu}_p W_{pg^\prime}^\nu W_{pg}^\nu
\label{mass-matrices-sub}
\end{equation}

Note that the definition of the sub-cycled mass matrix (eq.~(\ref{mass-matrices-sub})) and the sub-cycled hatted current (eq.~(\ref{hattedmoments-sub})) treats each sub-interval for each particle as if they were independent: the equations see each sub-interval as a particle. It is as if the system has $N_p N_\nu$ particles made by each particle for each sub-interval $\Delta_\nu$. This feature lends itself to a simpler computing implementation where each particle is spawned into $N_\nu$ treated as independent particles in the interpolation, mass matrix computation and current gathering step. A valuable approach in parallel  and vectorized computer architectures such as GPUs.

Regardless of how the positions for the particles during the sybcycling are chosen, if the current defined in eq. (\ref{currentECSIMfinal-sub}) is computed with  the mass matrices defined in eq. (\ref{mass-matrices-sub})  energy is conserved. In fact,  the term $\sum_g \overline{\bfJ}_{g} \cdot \overline{\bfE}_{g}$ is again identical when computed from the equations for the particles and for the fields.

\section{Simplification of the Mass Matrix: the limit of the Implicit Moment Method}
\label{simplification}

There is a specific case where the mass matrix formulation takes a much simplified form: in case of the nearest grid point (NGP) interpolation. In that case, the interpolation function $W_{pg}$ is simple: 1 for the nearest grid point, that we label as $g_p$, and 0 everywhere else: 
\begin{equation}
    W_{pg} = \delta_{g_pg}
\end{equation}
In this cases the mass matrix becomes:
\begin{equation}
M_{s,gg^\prime}^{ij} = \sum_p q_p {\alpha}^{ij,n}_p \delta_{g_pg^\prime} \delta_{g_pg}
\label{mass-matrices-ngp}
\end{equation}
When substituted in the expression for the current this leads to:
\begin{equation}
\overline{\bfJ}_{sg}=\widehat{\bJ}_{sg}+\frac{\beta_s }{V_g}\sum_p \sum_{g^\prime} q_p {\alpha}^{n}_p
 \bE_{g^\prime}^{n+\theta}  \delta_{g_pg^\prime} \delta_{g_pg}
\end{equation}
Given the properties of Kronecker's delta, the summation over $g^\prime$ can be done first:
\begin{equation}
    \sum_{g^\prime} \bE_{g^\prime}^{n+\theta}  \delta_{g_pg^\prime} = \bE_{g_p}^{n+\theta}
\end{equation}
and substituting:
\begin{equation}
\overline{\bfJ}_{sg}=\widehat{\bJ}_{sg}+\frac{\beta_s }{V_g}\sum_p  q_p {\alpha}^{n}_{g_p}
  \bE_{g_p}^{n+\theta}  \delta_{g_pg}
\end{equation}
where we have used the fact that the $\alpha$'s are computed using the magnetic fields of the nearest grid point, consistent with the NGP interpolation. 
Using again the properties of Kronecker's delta, the summation over the particles can be done directly:
\begin{equation}
\frac{1}{V_g}\sum_p  q_p = \rho_{sg}
\end{equation}
to obtain:
\begin{equation}
\overline{\bfJ}_{sg}=\widehat{\bJ}_{sg}+\beta_s  \rho_{sg} {\alpha}^{n}_g
  \bE_g^{n+\theta}  
  \label{current-ngp}
\end{equation}
This is the same expression that links the electric field and the current in the implicit moment method~\cite{riccijcp} used in Venus2D~\cite{brackbill-forslund}, in Celeste3D~\cite{lapenta05} and iPic3D~\cite{ipic3d}.
It is then worth considering this limit expression for the mass matrix formulation. Most often, the NGP cannot be used in practice for its well known excessive noise, but we can still use the expression \ref{current-ngp} even in presence of other interpolation schemes.

Eq. (\ref{current-ngp}) is exact only for the NGP scheme where it still leads to exact energy conservation. When, instead, other orders of interpolation are used, energy conservation is lost but it still provides a meaningful link between current and electric field: it is in fact the same used for decades by the implicit moment method. If eq. (\ref{current-ngp}) is used, ECsim becomes more similar to the implicit moment method but it still differs in one key aspect: the interpolations are computed using the position provided by the leap-frog algorithm for the particle position. This is known explicitly and does not require any iteration. In the implicit moment method, the mover requires to iterate between velocity update and position update using a predictor-corrector scheme. This is not required in ECsim where the particle position is known explicitly from the previous time step.

\section{Results}
\label{results}

\subsection{Effects of sub-cycling}

To check how the innovations described above perform in practice, we consider first the effect of sub-cycling. The goal is twofold. First, we want to verify that the energy is indeed conserved and sub-cycling does not break the energy conservation. Second, we test on a specific case how far one can push the sub-cycling before the physics deteriorates. The first task above is just a confirmation of the rigorously exact calculations above and it is merely a verification test for the code implemented here. The second task is only an illustration because the usefulness of sub-cycling is highly problem dependent. 

A two-stream instability is initiated by considering a domain of $L/d_{e}=2\pi$ divided into 64 cells with 10,000 particles. A mini-app is useful to study readily implementations in different computer architectures: it is intended merely as a test not as a real world problem. Space is normalized in electron skin depth, $d_e=c/\omega_{pe}$, and time in electron plasma frequency, $\omega_{pe}t$.
The ions are kept immobile while the electrons have a thermal speed of $v_{the}/c=0.02$ and the two beams have a net speed of $V_0/c=\pm 0.1$. The reference case without sub-cycling ($N_\nu=1$) 
has a CFL = $V_0\Delta t /\Delta x=0.1$. We then proceed to do 10 runs with $N_\nu=1-10$ and $\Delta t $ is increased by the same factor. That is we keep the particle $\Delta t$ fixed and increase the  $\Delta t$ of the field. As the number of sub-cycles is increased the $\Delta t$ increases but the total time remains the same: $\omega_{pe}t=50$. A perturbation of the particle velocity is added with mode number $m=5$:
\begin{equation}
    u_p=u_p+V_1 \sin\left(\frac{2 \pi m x_p}{L}\right);
\end{equation}
with an amplitude $V_1=v_0/10$. 

We use a MATLAB implementation that we test on a MAC OSX 10.14.6 with processor Intel Core i7 2,6GHz complemented with 16GB DDR3 1600 MHz memory. 

\begin{figure}
    \centering
    \includegraphics[height=.8\textheight]{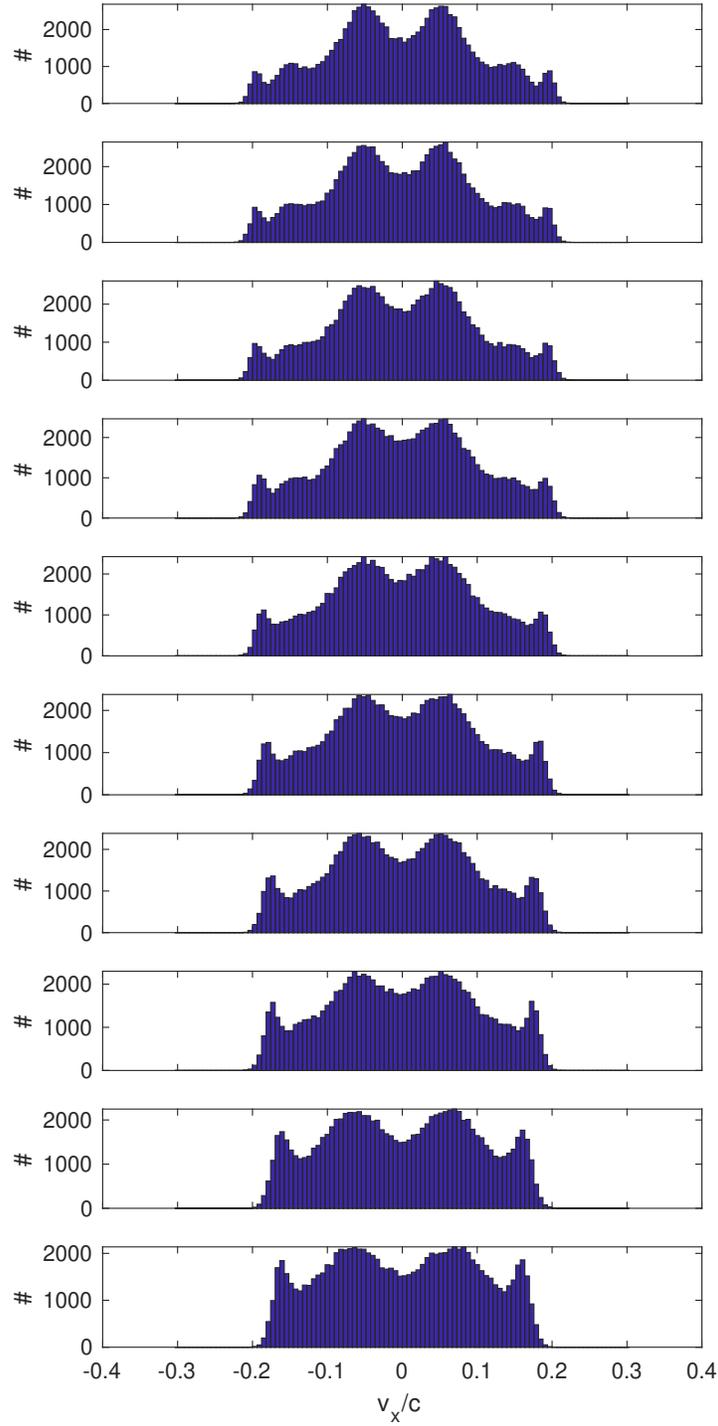}
    \caption{Particle velocity distribution at the end of 10 different simulations of same total duration varying the number of sub-cycles $N_\nu=1 - 10$ (from top, $N_\nu=1$, to bottom, $N_\nu=10$) and correspondingly increasing $\Delta t$ by the same factor.}
    \label{fig1}
\end{figure}

\begin{figure}
    \centering
    \includegraphics[height=.8\textheight]{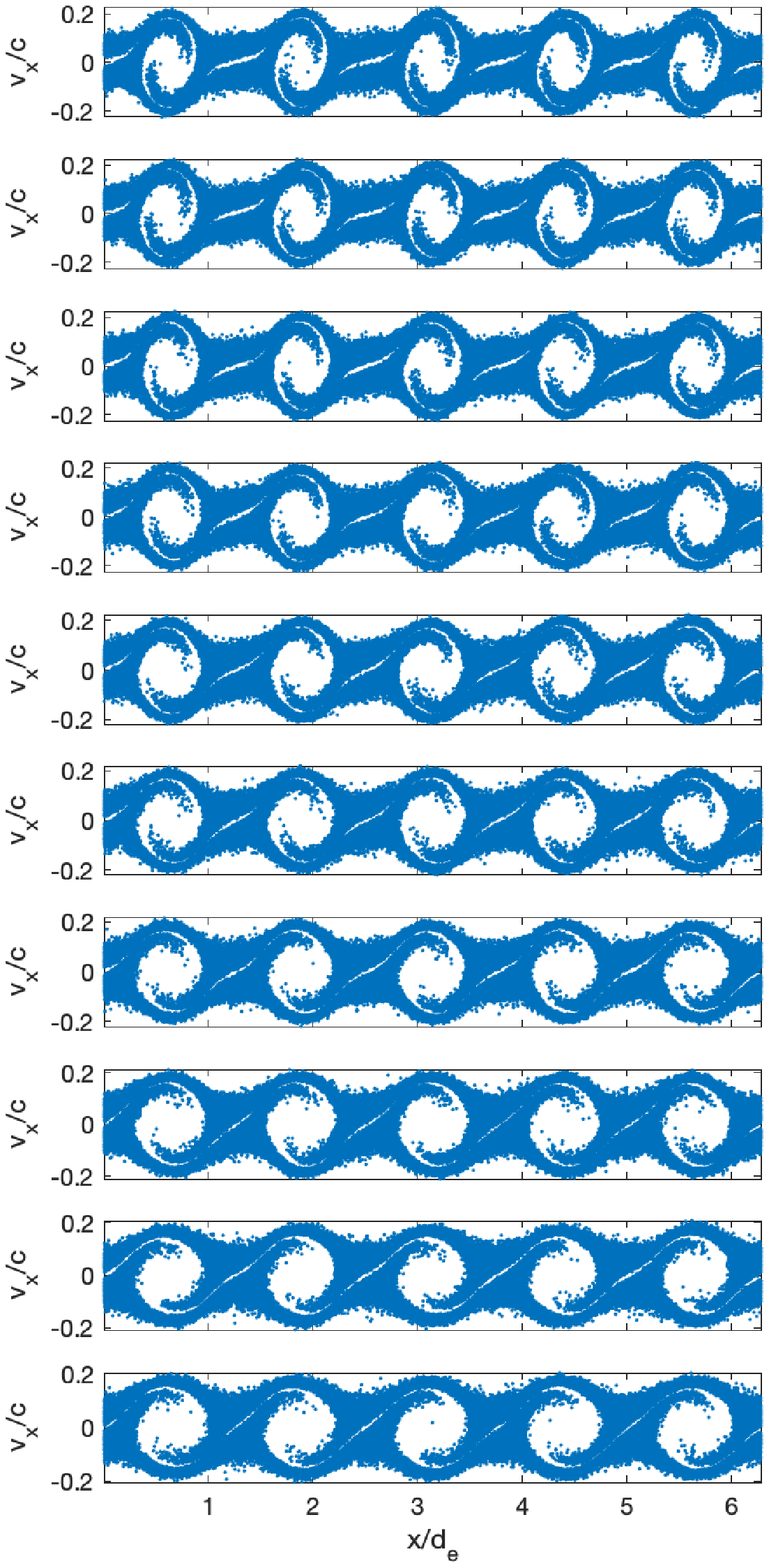}
    \caption{Scatter plot on the phase space for the particles  at the end of 10 different simulations of same total duration varying the number of sub-cycles $N_\nu=1 - 10$ (from top, $N_\nu=1$, to bottom, $N_\nu=10$) and correspondingly increasing $\Delta t$ by the same factor.}
    \label{fig2}
\end{figure}

The results of the study are summarized in Table ~\ref{tab1} and on Fig.~\ref{fig1}
-\ref{fig2}. 
\begin{table}
\centering
\begin{tabular}{||c|c|c|c||} \hline\hline $N_\nu$ & Time (s) & Seconds/ $\Delta t$ & $\Delta E$ \\ \hline 1.0 & 42.73 & 0.08395 & $1.75\,{10}^{-16}$\\ \hline 2.0 & 20.94 & 0.08244 & $1.1\,{10}^{-16}$\\ \hline 3.0 & 14.63 & 0.08657 & $1.015\,{10}^{-16}$\\\hline  4.0 & 10.88 & 0.08567 & $7.388\,{10}^{-17}$\\ \hline 5.0 & 8.97 & 0.08881 & $6.764\,{10}^{-17}$\\ \hline 6.0 & 7.36 & 0.08762 & $8.342\,{10}^{-17}$\\ \hline 7.0 & 6.46 & 0.08972 & $7.988\,{10}^{-17}$\\ \hline 8.0 & 5.74 & 0.09111 & $8.492\,{10}^{-17}$\\ \hline 9.0 & 5.27 & 0.09411 & $8.14\,{10}^{-17}$\\ \hline 10.0 & 4.79 & 0.0958 & $8.61\,{10}^{-17}$ \\ \hline\hline \end{tabular}
\caption{Effect of sub-cycling on a two-stream instability run. Varying the number of sub-cycles from 1 to 10, while correspondingly increasing by the same factor $\Delta t$, the duration of the simulation in seconds, the seconds spent per $\Delta t$ and the energy error are reported. }
\label{tab1}
\end{table}


Table ~\ref{tab1} reports the reduction of cost of the simulation as the number of sub-cycles is increased. The MATLAB implementation is well vectorized and efficient in handling the particle projections, so much so that as the number of sub-cycles is increased the cost per $\Delta t$ hardly increases at all. If one compares the cost of a time step without sub-cyclig (0.08395s) with that with 10 sub-cycles (0.0958s), the increase is minimal even though particles have been moved and projected to the moments and the construction of the mass matrix 10 times more. While this result is specific to the vectorization done via MATLAB, other forms of vectorizations are also possible on GPUs using OpenACC and CUDA, opening up a similar opportunity for this type of optminization also in other implementations.

As required by the theoretical derivations above, energy should be conserved in all cases. And this is indeed the case in Table ~\ref{tab1}, a confirmation that our MATLAB implementation is bug free.

But if sub-cycling  is beneficial in reducing the cost of the simulation, it introduces a degradation in the fidelity of the physics. Particles are moved with the same time step in each sub-cycle regardless of $N_\nu$ but as $N_\nu$ increases the overall time step $\Delta t $ for the field recalculation is increased. In a number of problem still it is beneficial to accept this compromise. Figure \ref{fig1} and  Fig. \ref{fig2} show the degradation of the results with $N_\nu$. Unfortunately the instability modeled is a complex non-linear process and the linear phase is not important: an easy quantitative metric of accuracy is not available. We start already from a significant perturbation. If we do not add that, several modes develop at the same time and still the linear phase is muddied by the interaction of many modes. There is no single metric that can easily summarize the quality of the evolution. We have to rely on qualitative visual comparison. For this reason we look carefully at the velocity space distribution  (Fig. \ref{fig1}) and phase space (Fig. \ref{fig2}). As $N_\nu$ is increased the correct physics is progressively lost.

The most characteristic  feature of the two stream instability is the formation of a flat top distribution, a distribution where for a range of velocities the distribution remains flat. This feature is prominent without sub-cycling, and it is still reasonably well represented  at higher sub-cycling numbers. However, the tail of higher energy particles is reduced when sub-cycling is increased, a reflection of the fact that the electric field responsible for particle acceleration~\cite{lapenta2007relaxation} is less accurately computed. These are non physical artifacts of excessive sub-cycling. 

The same conclusion is reached analysing phase space. In this case the most important feature is the formation of electron holes, regions of phase space depleted of electrons and in fact completely void of them. These features, observed also in experimental measurements, are distorted and expelled to the edge of the velocity range  as $N_\nu$ is increased.

sub-cycling  is compatible with the mass matrix formulation, retaining the property of exact energy conservation. It can be implemented very effectively, increasing only minimally the cost of the computational cycle, despite the increase in the number of particle operations, while reducing the number of time steps for the same total time. However, it must be used with care because not updating the fields after the particles are moved in a sub-cycle introduces physical errors even though energy is still exactly conserved.

\subsection{Effects of smoothing}
To test the effects of smoothing we consider another type of streaming instability: the transverse electromagnetic  instability driven by counter streaming beams \citep{doi:10.1063/1.1705933}. We consider again a 1D plasma ($x$ is the only spatial variable) but we consider now all three components of the particle velocity. The two counter streaming beams are  directed along $\mathbf{y}$. Using the speed of light $c$ for normalization,  we consider a case with the beams having $v_{th}/c=0.01$ counter streaming with speed $v_{0y}/c=0.2$. We add no initial perturbation and let the natural noise initiate the instability. The transverse electromagnetic streaming instability is well known \citep{doi:10.1063/1.1705933} and it is often used in the context of understanding how magnetic fields can be generated in the Universe \citep{lazar2009cosmological}, it is a form of dynamo. We use a setup similar to that reported in \citet{innocenti2011electron}. A characteristic of this instability is that it segregates in phase space particles with opposite signs of $v_y$, initially residing in the two different beams.

\begin{figure}
    \centering
    \begin{tabular}{cc}
    No smoothing & Smoothing\\
    a) $\omega_{pe} t=6.35$ & b) $\omega_{pe} t=6.35$ \\
    \includegraphics[width=.5\columnwidth]{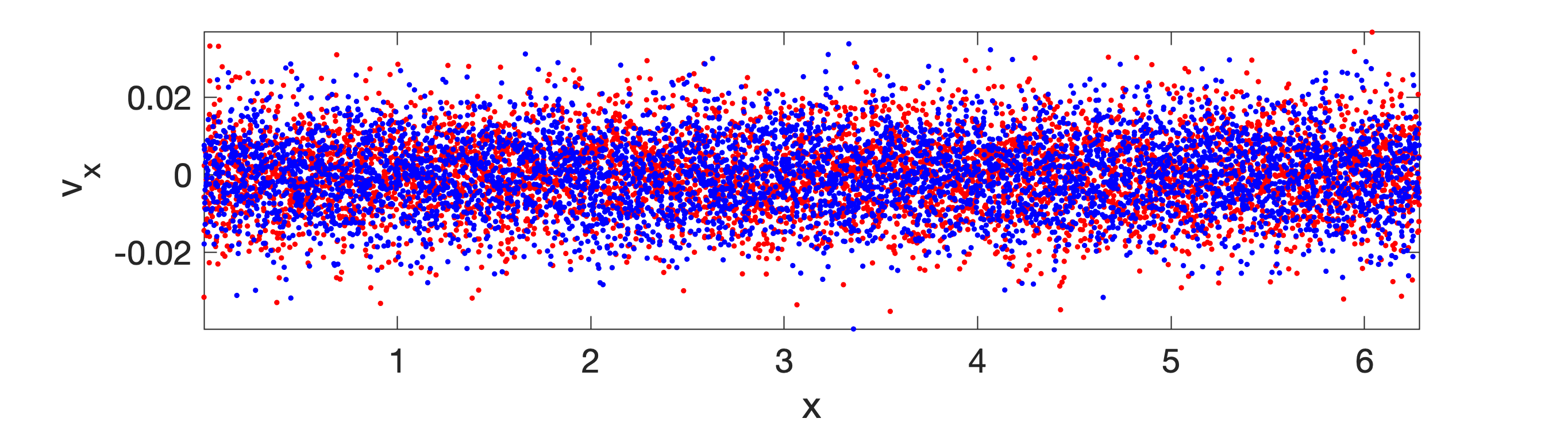} &  \includegraphics[width=.5\columnwidth]{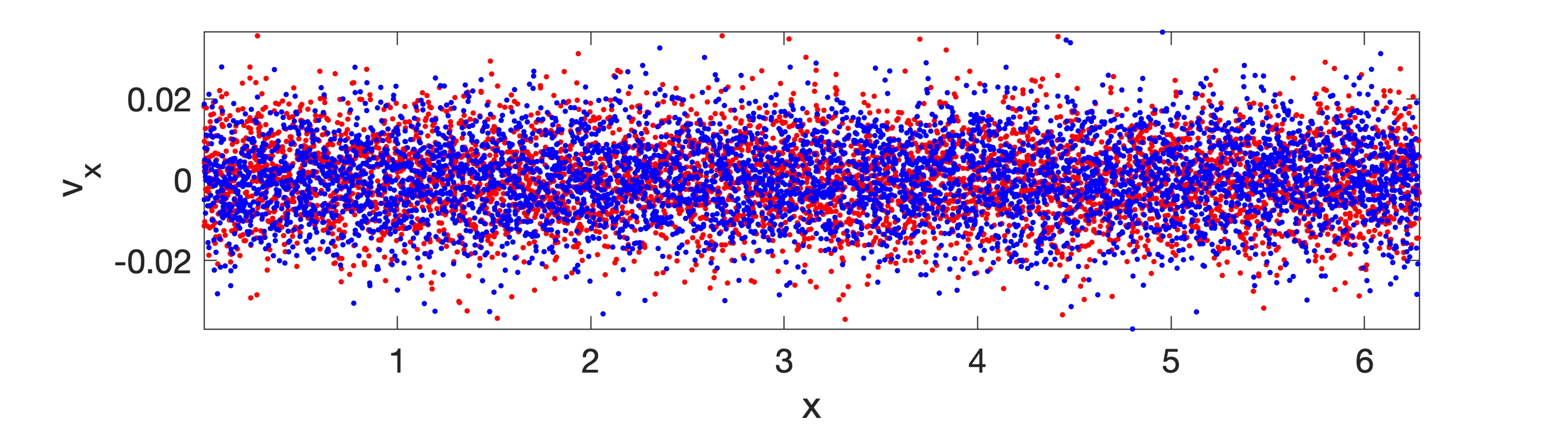} \\
        c) $\omega_{pe} t=25$ & d) $\omega_{pe} t=25$ \\
    \includegraphics[width=.5\columnwidth]{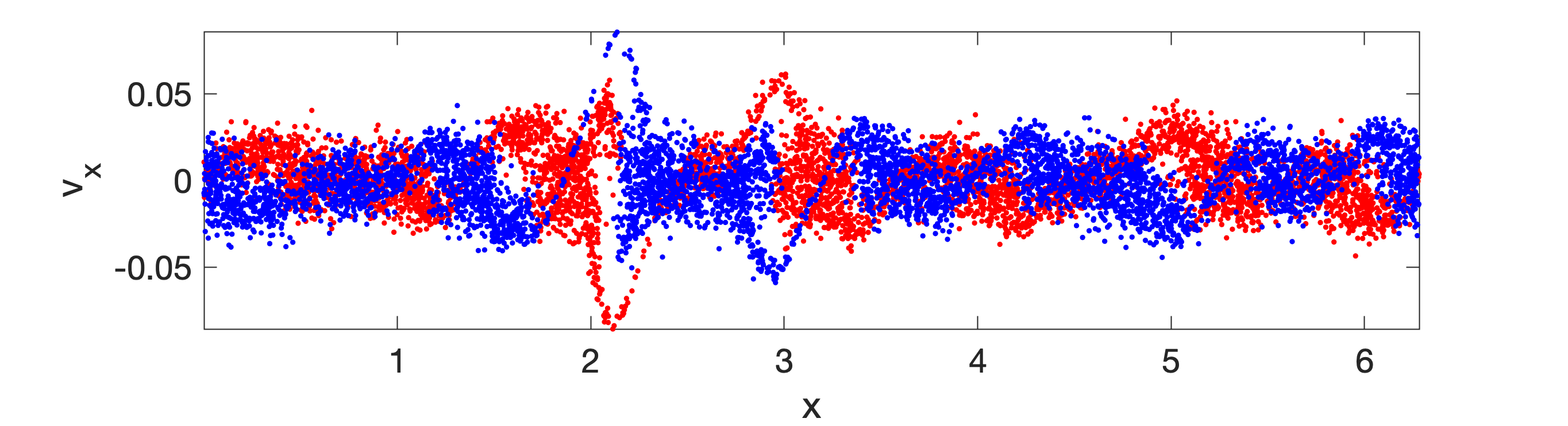} &  \includegraphics[width=.5\columnwidth]{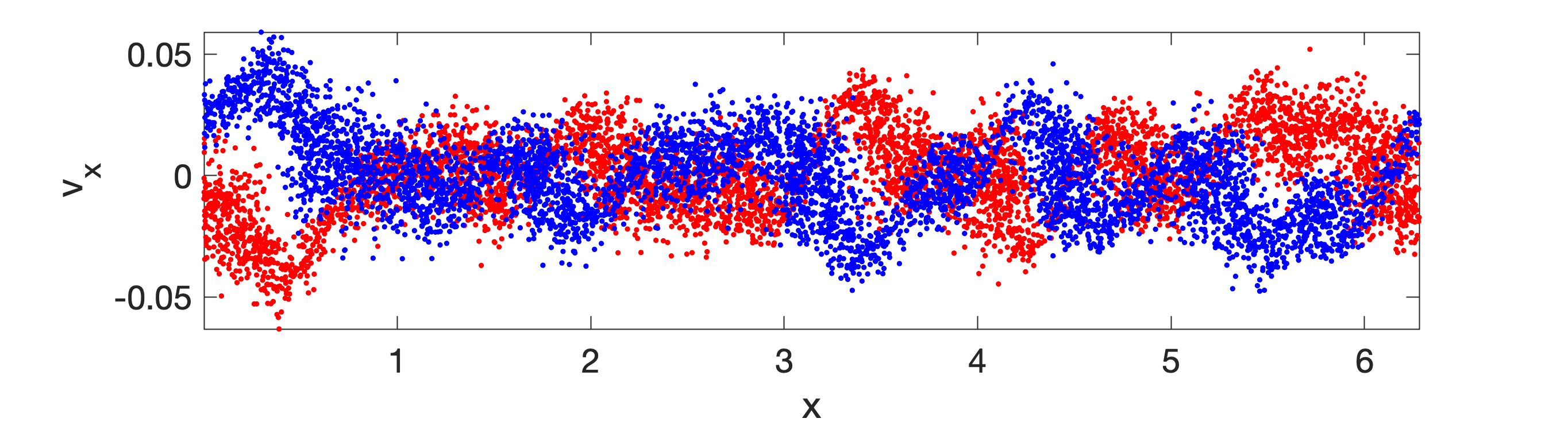} \\
            e) $\omega_{pe} t=43.75$ & f) $\omega_{pe} t=43.75$ \\
    \includegraphics[width=.5\columnwidth]{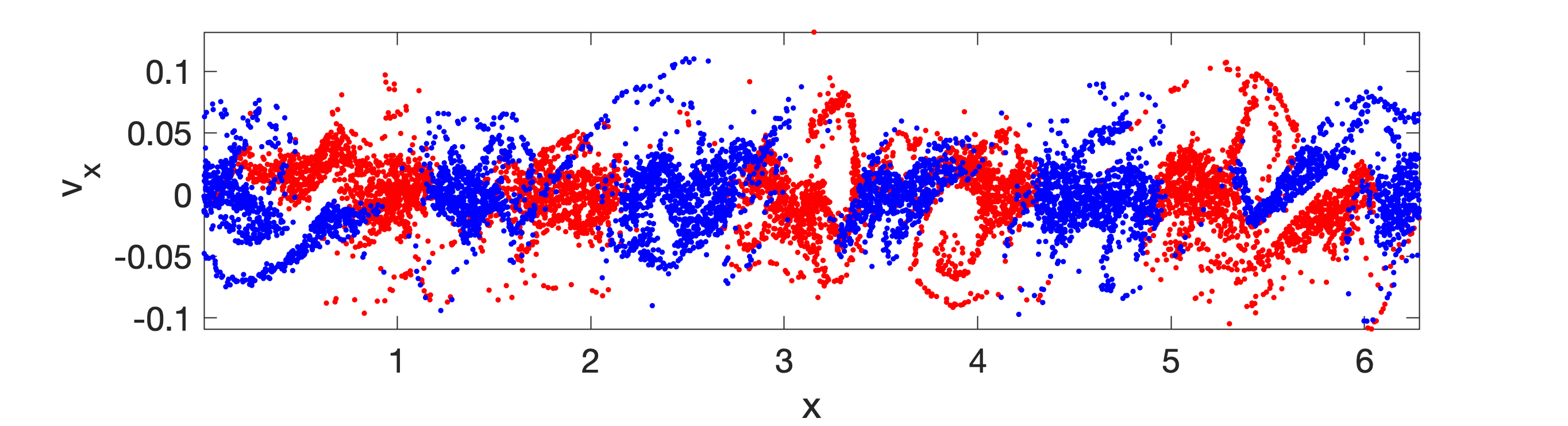} &  \includegraphics[width=.5\columnwidth]{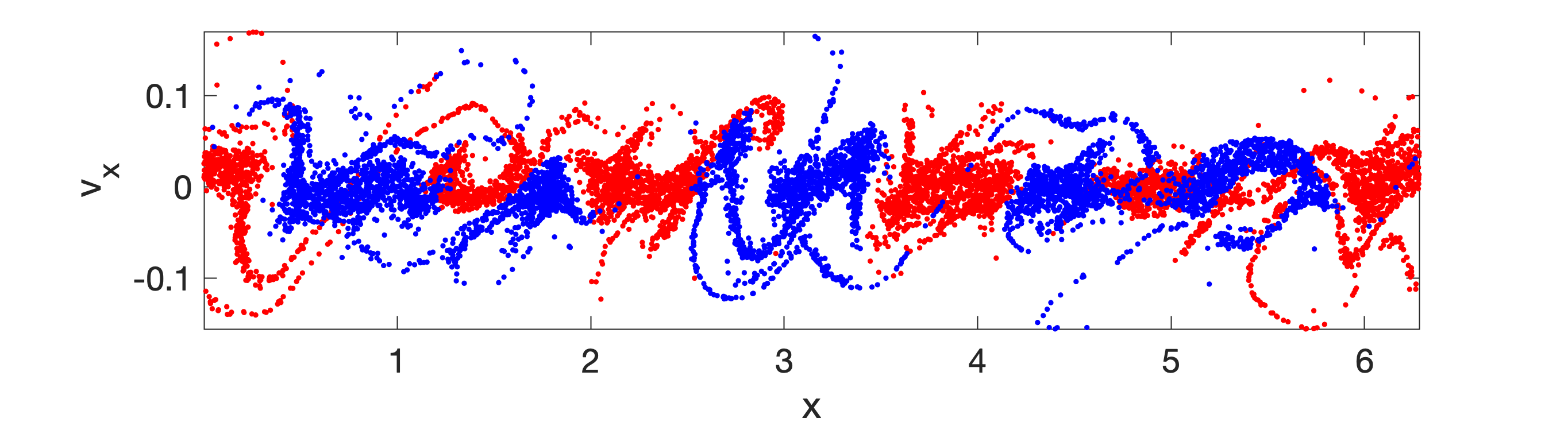} \\
                g) $\omega_{pe} t=62.5$ & h) $\omega_{pe} t=62.5$ \\
    \includegraphics[width=.5\columnwidth]{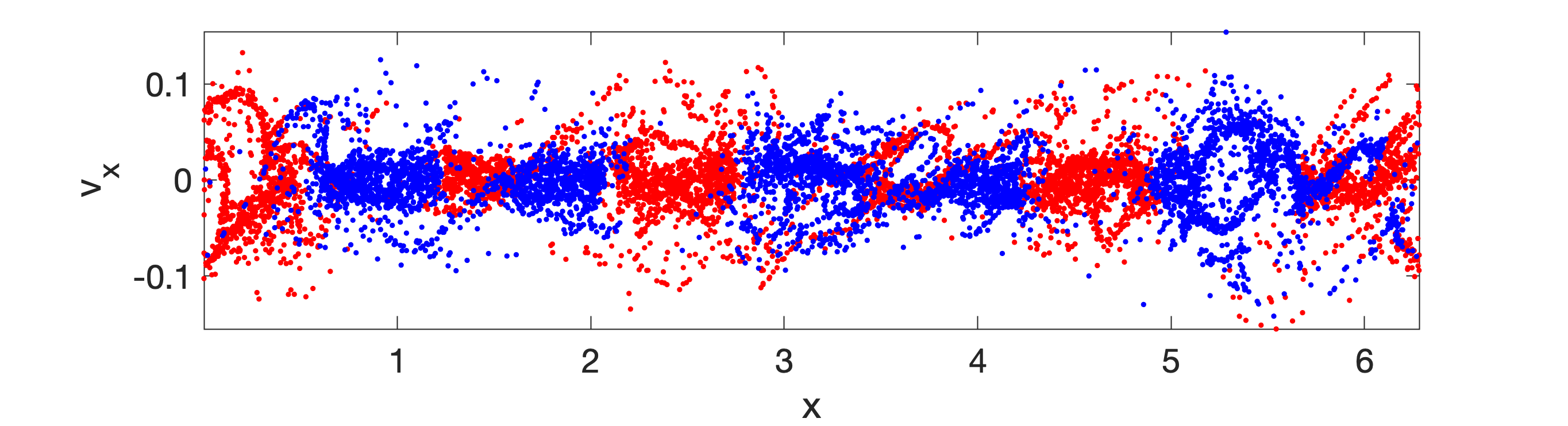} &  \includegraphics[width=.5\columnwidth]{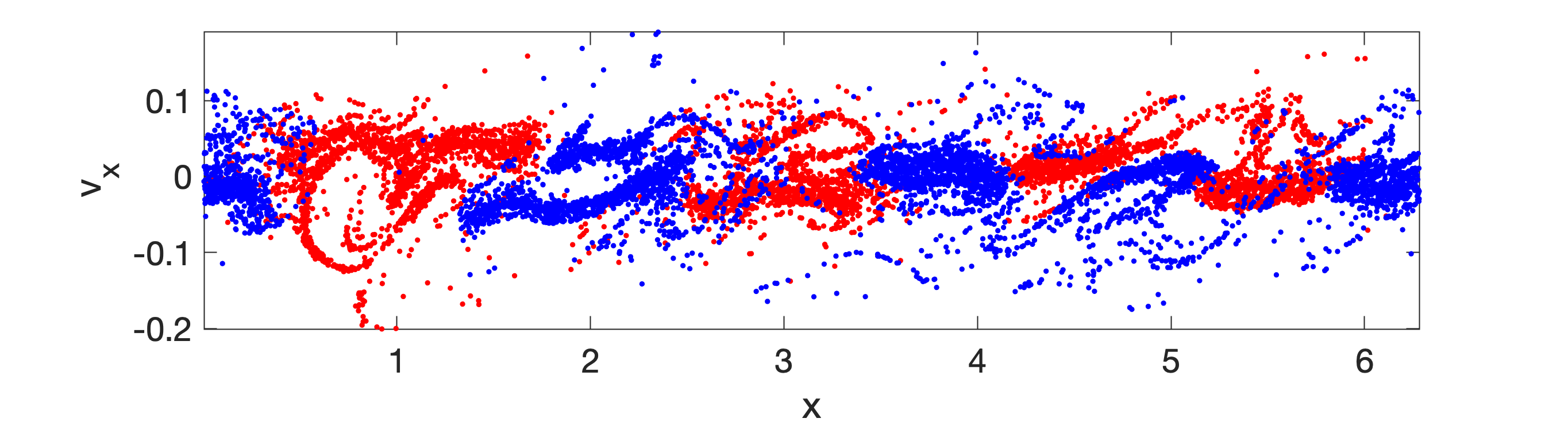} \\
    \end{tabular}
    \caption{Evolution of the phase space section  $(x,v_x)$ in the transverse counter streaming  instability. The red particles have $v_y<0$ and the blue particles $v_y>0$. Three times are shown and the left (right) panels report the non smoothed (smoothed) run. }
    \label{smooth_ps}
\end{figure}

Figure \ref{smooth_ps} shows the evolution of the phase space cross section $(x,v_x)$: the red particles have $v_y<0$ and the blue particles $v_y>0$. The evolution initially retains their separation but in time phase space mixing takes over. More details of the physics of this simulation can be found in \citet{innocenti2011electron}.
We are focusing here on comparing the normal ECsim case without smoothing with one where the smoothing kernel $\mathbb{S}=[1/4,1/2,1/4]$ is applied by convolution 3 times. 
Figure \ref{smooth_ps}  compares the smoothed and non smoothed case. Smoothing is not altering in any profound ways the evolution but it affects it. Since the simulation is started from its natural noise, no two simulations will be identical and of course the smoothed simulation will have less noise. The location of the islands formed in phase space is not the same but the overall features are similar. 

\begin{figure}
    \centering
    \begin{tabular}{cc}
    No smoothing & Smoothing\\
    \includegraphics[width=.5\columnwidth]{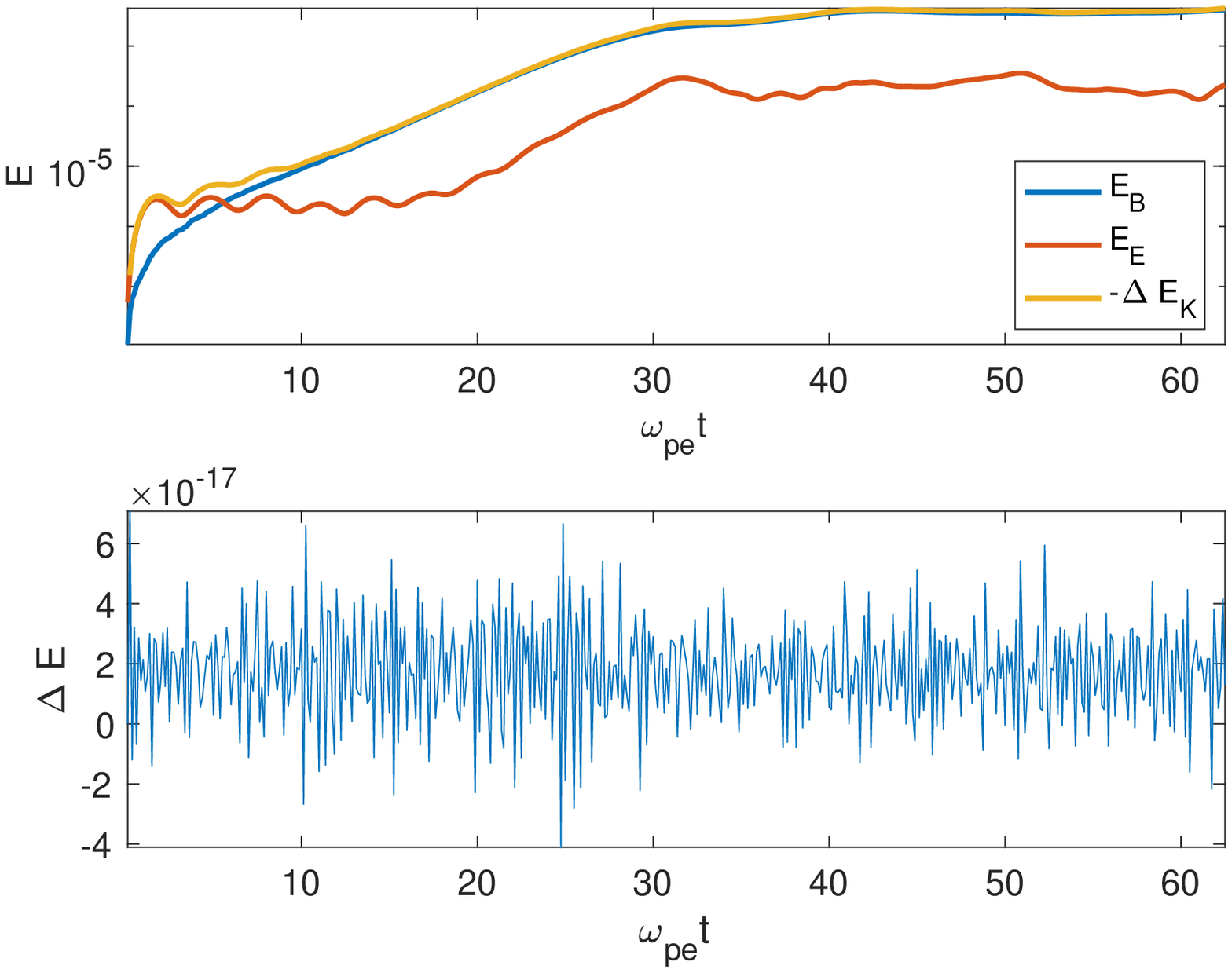} &  \includegraphics[width=.5\columnwidth]{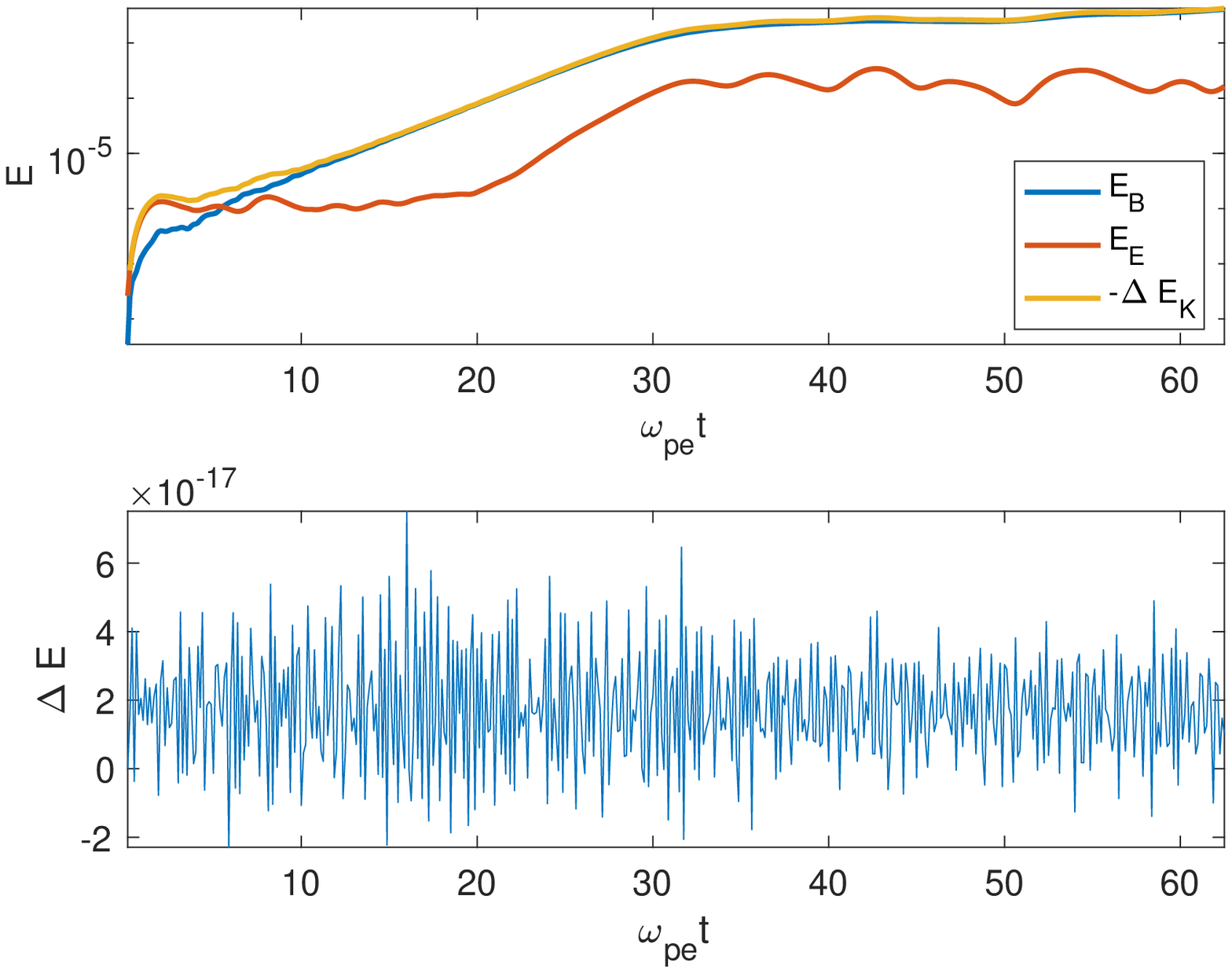} 
    \end{tabular}
    \caption{Evolution of the energy in the transverse counter streaming  instability. The left panels show the case without smoothing and the right panels with smoothing. The tope panel show the exchange between kinetic and magnetic energy with a minority contribution going to the electric field energy. The bottom panel shows the total energy conservation.  }
    \label{smooth_energy}
\end{figure}

The energy exchange is also similar in the smoothed and non smoothed run, as shown in Fig.~\ref{smooth_energy}. In particular the kinetic energy is lost primarily to produce magnetic energy. The transverse counter streaming instability is a form of magnetic dynamo that spontaneously creates a magnetic field by using the kinetic energy of the counter streaming beams. The total energy is conserved to machine precision in both runs. Smoothing, as shown in Sect.~\ref{smoothing} theoretically, indeed conserves energy exactly. 

\begin{figure}
    \centering
    \begin{tabular}{ccc}
    \includegraphics[width=.3\columnwidth]{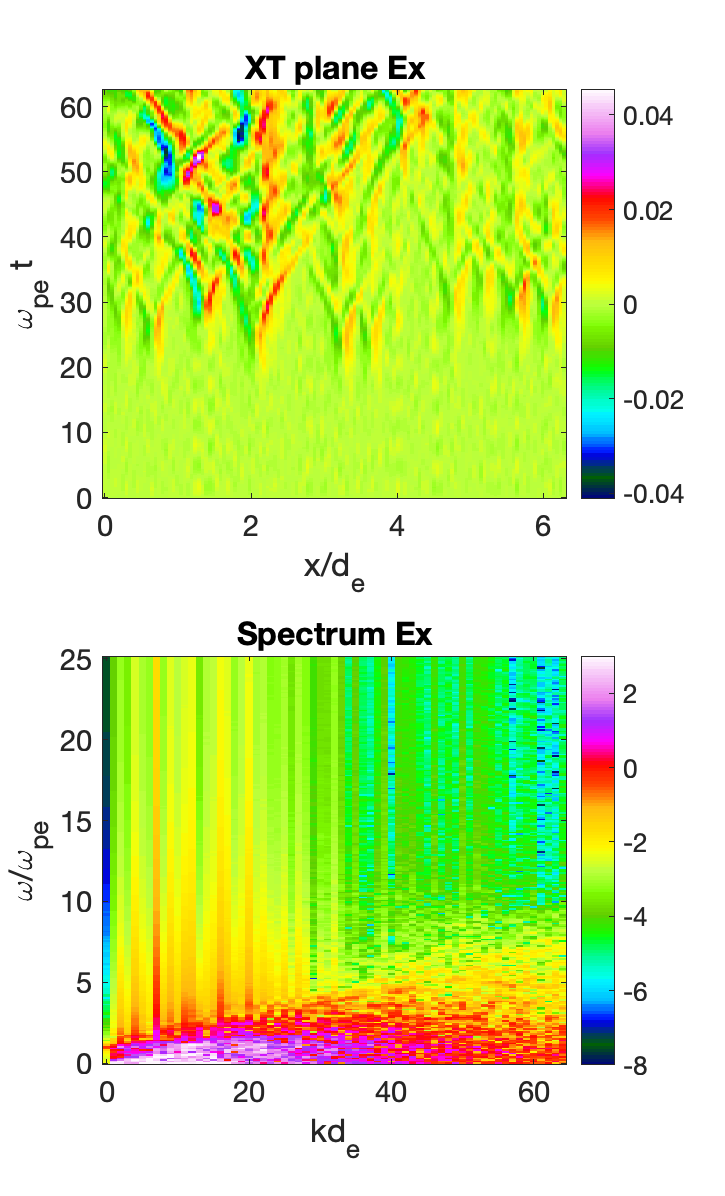} &  \includegraphics[width=.3\columnwidth]{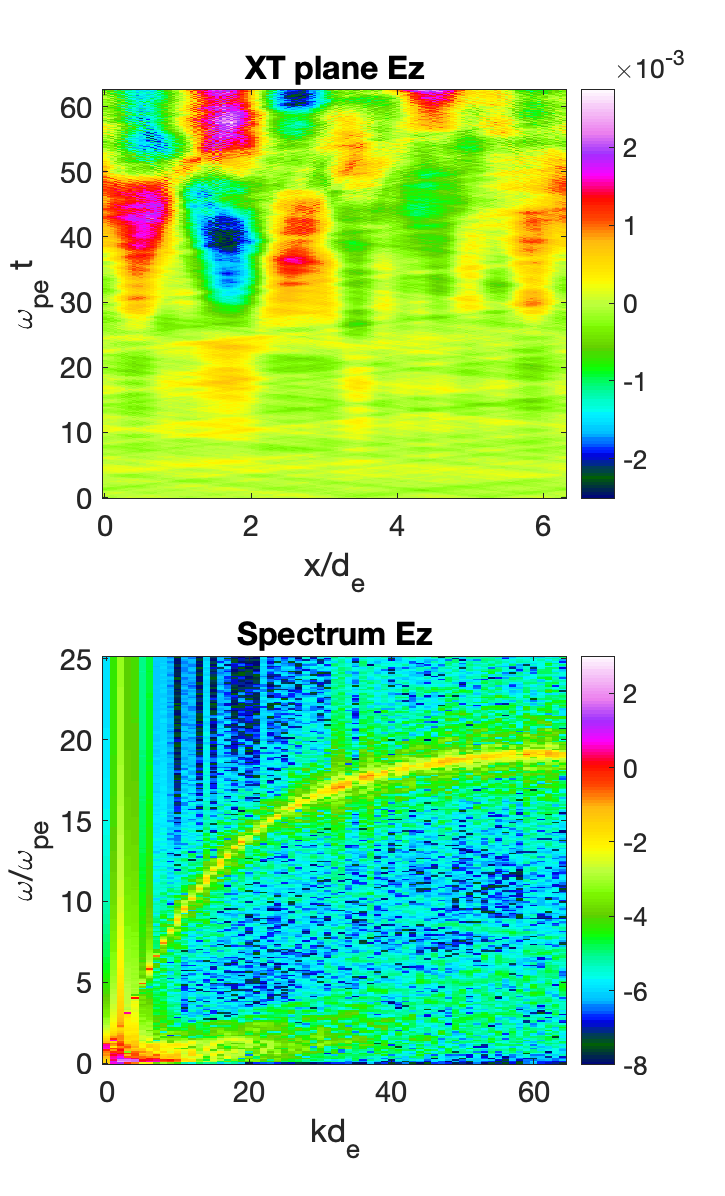}  & \includegraphics[width=.3\columnwidth]{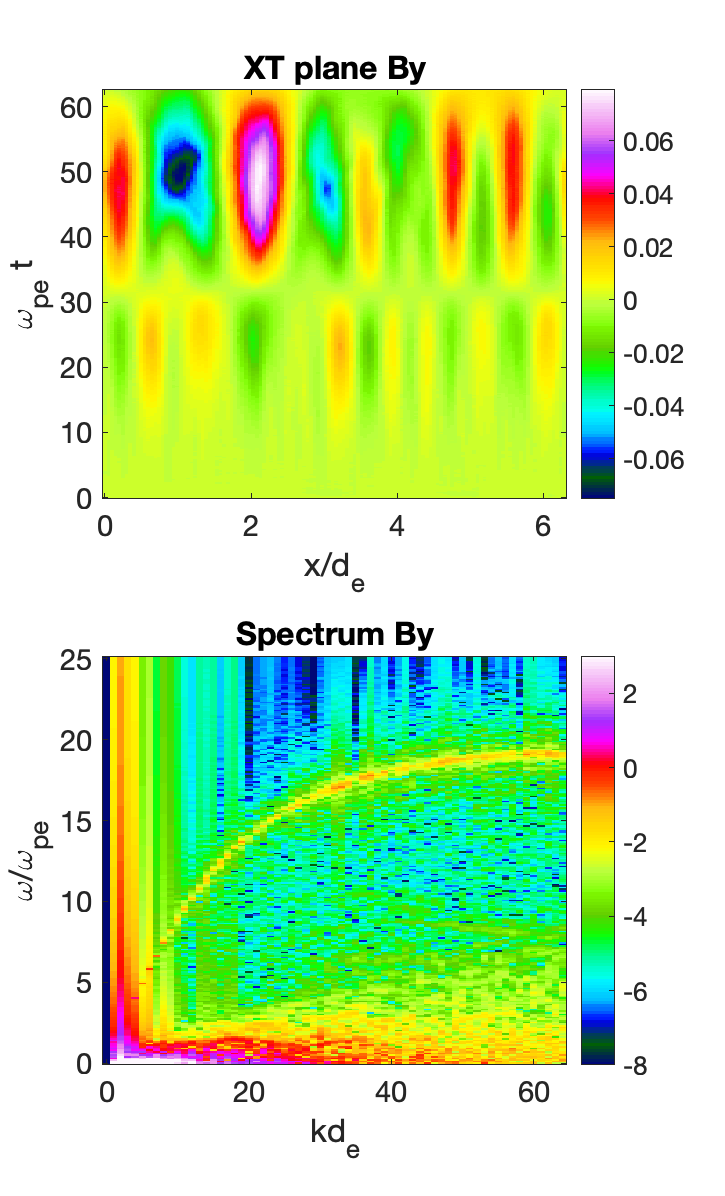} 
    \end{tabular}
    \caption{Transverse counter streaming  instability in absence of smoothing. Three fields are shown from left to right: $E_x$, $E_z$ and $B_y$. The top row shows the spatiotemporal plane $(x,t)$ and the bottom row the spectral plane $(k,\omega)$. }
    \label{smooth2}
\end{figure}

\begin{figure}
    \centering
    \begin{tabular}{ccc}
    \includegraphics[width=.3\columnwidth]{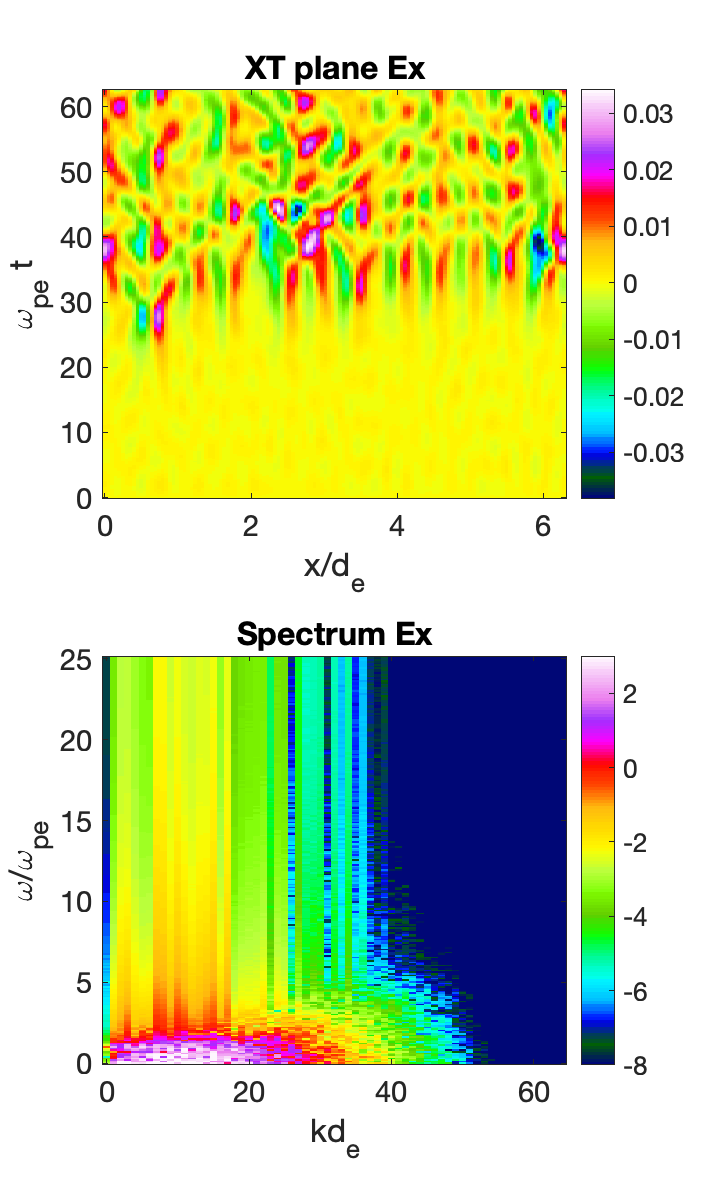} &  \includegraphics[width=.3\columnwidth]{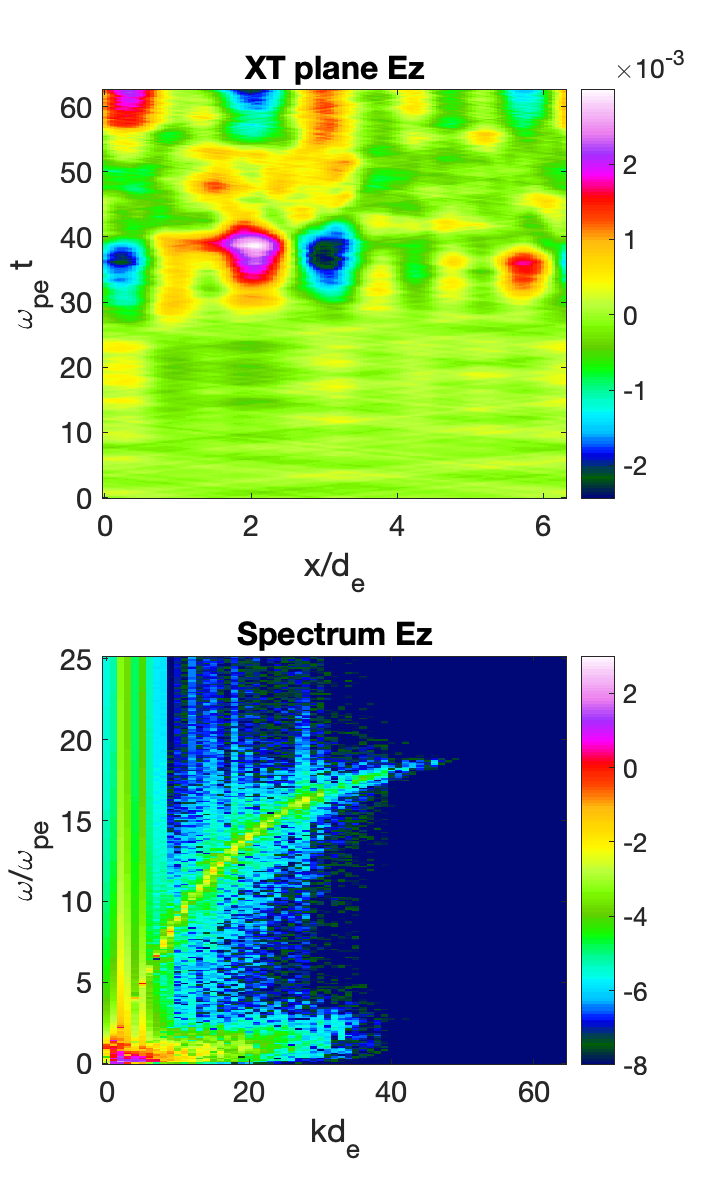}  & \includegraphics[width=.3\columnwidth]{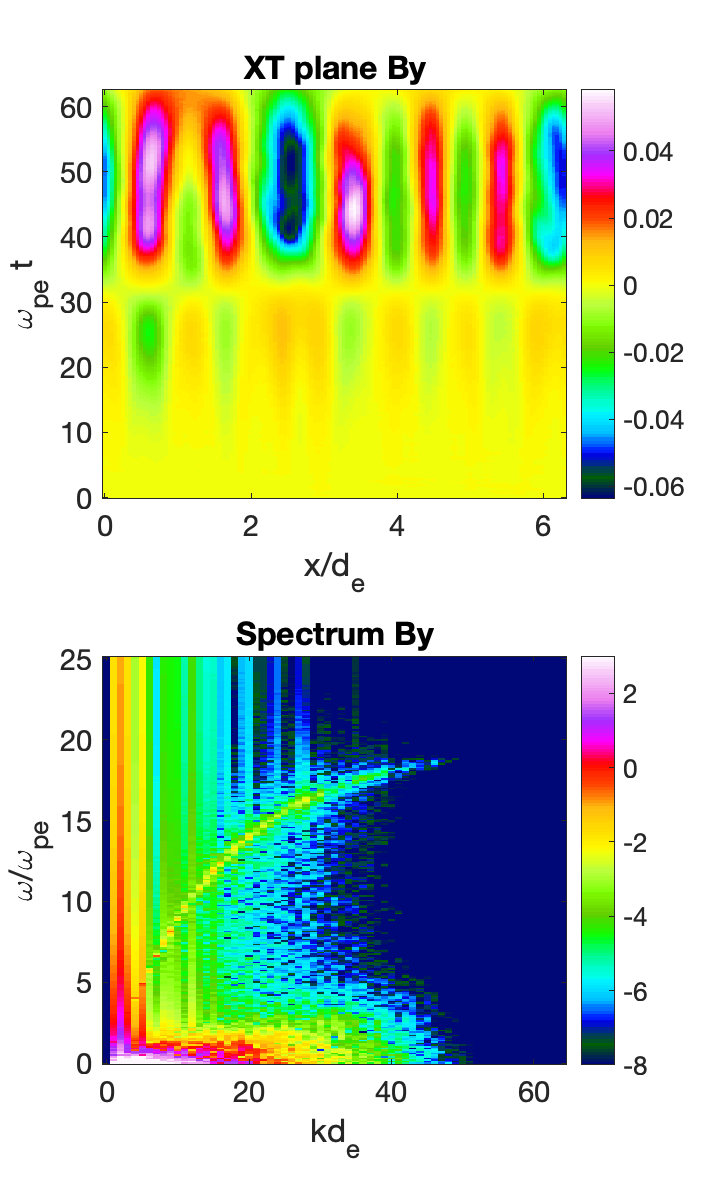} 
    \end{tabular}
\caption{Transverse counter streaming instability in presence of smoothing. Three fields are shown from left to right: $E_x$, $E_z$ and $B_y$. The top row shows the spatiotemporal plane $(x,t)$ and the bottom row the spectral plane $(k,\omega)$. }
    \label{smooth1}
\end{figure}

The main effect of smoothing is to eliminate the high frequency part of the spectrum. Figure  \ref{smooth2} and \ref{smooth1} compares the $k-\omega$ spectrum of selected fields. The noise at high values of $k$ is reduced. The spectrum identifies the low $k$ part of the spectrum as dominant. The characteristic arch of the electromagnetic waves (light waves) is also prominent.

%
%

\section{Discussion}
\label{discussion} 

The ECsim method allows within a semi-implicit approach to make PIC simulation that conserve energy exactly. This feature besides having its own intrinsic value, leads also to improved stability allowing, for example, to consider realistic plasma conditions in the heliosphere, including the colder solar wind where the Debye length is very small compared with the scales of interest. 

We report here three new developments. 

First, we introduced a method for smoothing the electric field in an algorithm that preserves energy conservation. When it can be avoided, smoothing should be avoided, but when there is a need for it, the algorithm presented here achieves smoothing without breaking energy conservation. This can be for example the case when excessive noise alters the correct transport properties of a plasma, an issue of great importance for example in fusion energy studies~\citep{nevins2005discrete}. 

Second, we introduced a method for computing the mass matrix in presence of particle sub-cycling. Sub-cycling can be useful in a number of situations. Most often if it can be avoided, it should be avoided because particles and fields move together, the frozen-in condition being a cardinal property of plasmas. However, frozen in is valid at the large scales typical of MHD. There are many situations where the fields evolve slowly while particles move quickly. For example in gyro motion. Sub-cycling can be used together with gyroaveraging. The method presented here allows to use sub-cycling while constructing a mass matrix that continues to preserve the exact energy conservation. 

Finally, we discussed a limit case when the mass matrix calculation becomes especially simple and show that in this limit the plasma particle response in ECsim method becomes identical to that of the implicit moment method (IMM). This has two implications. First, it allows to use the same code either in the full energy conserving ECsim mode or in the less expensive IMM mode. Comparisons can then be made more readily.  Second, the derivation clarifies the theoretical links between ECsim and IMM, revealing in what limit the two become identical.

These three new steps have both a theoretical significance and practical value, bringing understanding of the properties of ECsim and broadening its range of applications. 
%

%
%
\vspace{6pt}

\acknowledgments 
The work reported received funding from the KULeuven Bijzonder Onderzoeksfonds (BOF) under the C1 project TRACESpace and from the European Union project DEEP-SEA (grant agreement 955606).


\end{document}